\documentclass[english,prb,aps,twocolumn,10pt]{revtex4-2}

\usepackage{amsmath}
\usepackage{amssymb}
\usepackage{graphicx}
\usepackage{babel}
\usepackage{float}
\usepackage{array}
\usepackage{verbatim}
\usepackage[colorlinks=true, pdfstartview=FitV, linkcolor=blue, citecolor=blue, urlcolor=blue]{hyperref} 

\def\eg       {{\it e.g.}}

\newcommand{\captionstyle}{\normalfont} 

\newcommand{\ee}[1]{\cdot10^{#1}}
\newcommand{\mr}[1]{\mathrm{#1}}
\newcommand{\unit}[1]{\,\mathrm{#1}}
\newcommand{\um}{\,\mathrm{\mu m}}
\newcommand{\nm}{\,{\rm nm}}
\newcommand{\us}{\,\mathrm{\mu s}}
\newcommand{\uT}{\,\mu{\rm T}}
\newcommand{\mT}{\,{\rm mT}}
\newcommand{\MHz}{\,{\rm MHz}}

\newcommand{\uA}{\,\mu{\rm A}}

\newcommand{\rtHz}{\sqrt{\mr{Hz}}}

\newcommand{\degree}{^\circ}
\newcommand{\ye}{\gamma_\mr{e}}

\newcommand{\Bac}{B_\mr{ac}}
\newcommand{\Bmax}{B_\mr{max}}

\newcommand{\Crefa}{C_\mr{ref}^0}
\newcommand{\Crefb}{C_\mr{ref}^1}

\newcommand{\Isd}{I_\mr{SD}}
\newcommand{\Ioffset}{I_\mr{offset}}

\newcommand{\vecJ}{\bf J}

\newcommand{\Vbg}{V_\mr{BG}}
\newcommand{\Vsd}{V_\mr{SD}}

\newcommand{\phiw}{\phi_\mr{wrapped}}

\begin{document}

\title{Imaging of sub-$\mu$A currents in bilayer graphene using a scanning diamond magnetometer}

\author{M.~L.~Palm$^{1,\dagger}$, W.~S.~Huxter$^{1,\dagger}$, P.~Welter$^1$, S.~Ernst$^1$, P.~J.~Scheidegger$^1$, S.~Diesch$^1$, K.~Chang$^{1,\ddagger}$, P.~Rickhaus$^{1,\S}$, T. Taniguchi$^{2}$, K. Wantanabe$^{3}$, K.~Ensslin$^{1,4}$, and C.~L.~Degen$^{1,4}$}

\affiliation{$^1$Department of Physics, ETH Zurich, Otto Stern Weg 1, 8093 Zurich, Switzerland;}
\affiliation{$^2$International Center for Materials Nanoarchitectonics, National Institute for Materials Science, 1-1 Namiki, Tsukuba 305-0044, Japan;}
\affiliation{$^3$Research Center for Functional Materials, National Institute for Materials Science, 1-1 Namiki, Tsukuba 305-0044, Japan;}
\affiliation{$^4$Quantum Center, ETH Zurich, 8093 Zurich, Switzerland.}

\email{degenc@ethz.ch}
\thanks{$^\dagger$These authors contributed equally.}
\thanks{$^\ddagger$Present address: Aeva Inc., 555 Ellis St., Mountain View, CA 94043, USA.}
\thanks{$^\S$Present address: Qnami AG, Hofackerstrasse 40B, 4132 Muttenz, Switzerland.}

\begin{abstract}
Nanoscale electronic transport gives rise to a number of intriguing physical phenomena that are accompanied by distinct spatial patterns of current flow.  Here, we report on sensitive magnetic imaging of two-dimensional current distributions in bilayer graphene at room temperature.  By combining dynamical modulation of the source-drain current with ac quantum sensing of a nitrogen-vacancy center in a diamond probe, we acquire magnetic field and current density maps with excellent sensitivities of 4.6\,nT and 20\,nA/$\mu$m, respectively.  The spatial resolution is 50-100\,nm.  We further introduce a set of methods for increasing the technique's dynamic range and for mitigating undesired back-action of magnetometry operation on the electronic transport.  Current density maps reveal local variations in the flow pattern and global tuning of current flow via the back-gate potential.  No signatures of hydrodynamic transport are observed.  Our experiments demonstrate the feasibility for imaging subtle features of nanoscale transport in two-dimensional materials and conductors.
\end{abstract}

\date{\today}

\maketitle


%
Electronic transport in nanostructures and thin films shows a rich variety of physical effects that have been fundamental to the development of modern electronics and communication devices. The standard method for investigating electronic transport -- resistance measurements -- does not provide detailed information on the nanoscale current distribution in such structures. The lack of spatial information is unfortunate, because the current distribution plays a key role in many intriguing physical phenomena. Having a technique that could simply look at nanoscale current flow would be immensely valuable.

Over the recent past, researchers have made significant progress in the sub-$\um$ imaging of nanoscale transport phenomena using scanning probe techniques \cite{marchiori21}.  For example, scanning gate microscopy has been applied to image branched flow \cite{topinka01}, universal conductance fluctuations \cite{berezovsky10}, beam collimation \cite{bhandari18} and viscous electron flow \cite{braem18}.  Scanning SQuID-on-tip microscopy has been used for the thermal imaging of dissipation \cite{halbertal16}, the magnetic imaging of persistent edge currents \cite{uri20nphys} and twist-angle disorder \cite{uri20nature} in graphene devices.  Scanning single-electron transistors have demonstrated simultaneous mapping of electrostatic potential and current to visualize ballistic and hydrodynamic electron flow \cite{sulpizio19,ella19}.  Scanning diamond magnetometers based on nitrogen-vacancy (NV) centers have been used to record current profiles in Ohmic and hydrodynamic transport regimes in graphene \cite{jenkins20,ku20,lee21} and semi-metals \cite{vool21}.  Overall, these techniques have opened an exciting avenue for imaging nanoscale transport phenomena in real space.

Recent demonstrations with scanning diamond magnetometers applied fairly large source-drain currents, of order of a few $\uA$ \cite{chang17, ku20, vool21} to a few tens of $\uA$ \cite{chang17, jenkins20,vool21, ku20, lee21}, because of sensitivity limitations.  Although some phenomena can be observed under these conditions, high current densities are in general undesirable, as they can cause, \eg, a heating of the electron gas \cite{baker12} or non-linearities due to large source-drain potentials \cite{kouwenhoven89}.  Moreover, spatial features of interest often only amount to few-percent changes in the total current density, leading to challenges in background suppression.  Another concern is the influence of the probe tip \cite{eriksson96, berezovsky10}, the optical readout \cite{lee08nnano, cao16, ju14} and the microwave spin manipulation \cite{chang17} on the transport properties.  All of these issues provide strong motivation for further improving the sensitivity of the technique and exploring methods for mitigating undesired stray effects.

%
\begin{figure*}[t!]
	\includegraphics[width=0.98\textwidth]{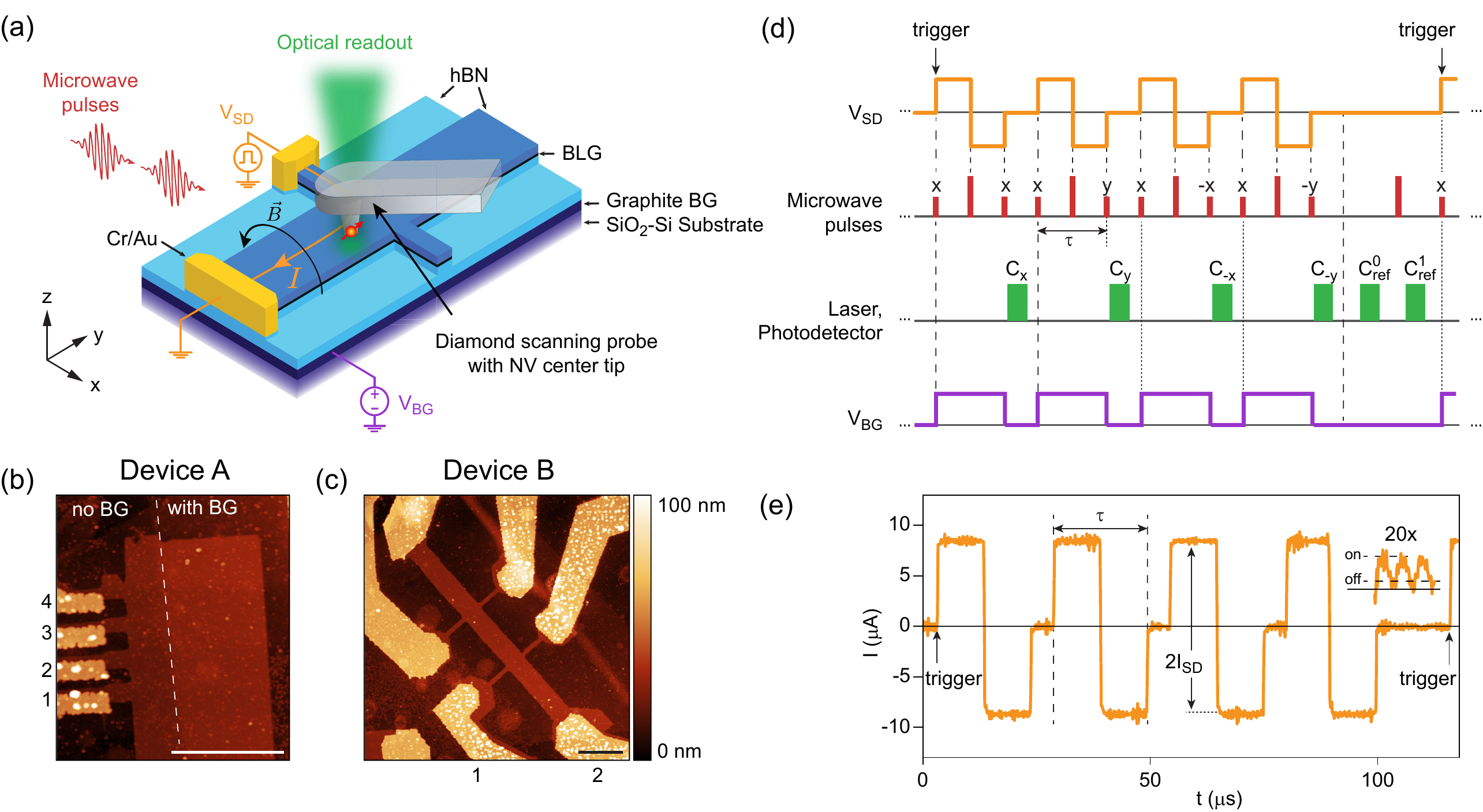}
	\caption{\captionstyle
	{\bf Schematic of the current imaging experiment}. 
	  (a) We use a nitrogen-vacancy center (red) in a diamond tip (gray) to image the magnetic stray field appearing above a current-carrying graphene device.  Microwave (dark red) and laser pulses (green) are used to manipulate and read out the spin state of the NV center.  The device consists of a bilayer graphene (BLG) sheet encapsulated in hexagonal boron nitride (hBN, 11\,nm top and 27\,nm bottom thickness) that sits on top of a graphite back-gate (BG). $\Vsd$ and $\Vbg$ are source-drain and back-gate voltages, respectively.
		(b,c) Atomic force microscopy images of the two devices used in this study. Note that device A is only partially covered by the back gate (white dashed line). Electrical contacts are numbered. Scale bars are $2\unit{\um}$.
		(d) Measurement protocol.  We modulate the source-drain voltage $\Vsd$ (orange), microwave power (red), laser power (green) and back-gate voltage $\Vbg$ (purple) using an arbitrary waveform generator.  Tall microwave pulses are $\pi$ rotations and short pulses are $\pi/2$ rotations. Labels $x,y,-x,-y$ indicate the pulse phase $\Phi$. $\tau$ is the phase accumulation time of the dynamical decoupling sequence (here a spin echo).  $\Crefa$ and $\Crefb$ are reference PL intensities of the two spin states. Triggers indicate the start of a measurement cycle.
		(e) Measured source-drain current during the experimental protocol (d). The inset shows the effect of laser pulses (on,off) when the tip is positioned near one of the injection points (10 point moving-average filter applied).  $2\Isd$ is the peak-to-peak amplitude and $\Ioffset$ is the dc offset.
	}
	\label{fig1}
\end{figure*}
%

In this work, we present advances to the sensitive and non-invasive imaging of current flow in two-dimensional materials using scanning diamond magnetometry.  Our samples are bilayer graphene (BLG) devices encapsulated in hexagonal boron nitride.  We demonstrate coherent detection of modulated (kHz-MHz) currents with sub-$\uA$ sensitivity and introduce a Bayesian quantum-phase unwrapping method for resolving small current density variations on top of large background currents.  We also investigate and mitigate the influence of the scanning tip, laser and microwave pulses on the transport properties.
We analyze current density maps for spatial variations in conductivity and show that the flow pattern can be deliberately changed by adjusting the carrier density via the back-gate potential.  Finally, opposed to recent imaging of monolayer graphene (MLG) \cite{ku20,jenkins20}, we observe no signatures of hydrodynamic transport for our bilayer graphene devices at room temperature.



\textit{Setup and devices -- }
A schematic of our experiment is shown in Fig.~\ref{fig1}a.  We study current flow in patterned graphene devices by recording magnetic field maps above the surface using a diamond scanning probe with an NV center tip \cite{degen08apl,balasubramanian08}.  Devices are fabricated from a single bilayer graphene sheet that is encapsulated between two layers of hexagonal boron nitride using mechanical exfoliation and stacking in a dry transfer process \cite{wang13,zomer14}.  The van-der-Waals stack is located on top of a 4-nm thick graphite flake acting as a back-gate.  The final stack is annealed, electrically contacted (Cr/Au) \cite{wang13} and patterned through e-beam lithography and reactive ion etching.
Two device geometries are used in this study: Device A is named the ``four-terminal device'' and is only partially covered by the back gate (Fig. \ref{fig1}(b)). Device B has a Hall-bar geometry and is fully covered by the back gate (Fig. \ref{fig1}(c)).  Conventional transport measurements at ambient conditions on the Hall bar (width: $0.8\unit{\um}$, side contact separation: $3\unit{\um}$) yield Hall mobilities for electrons (holes) of $\mu \approx 3.3\ee{4}\unit{cm^2/(Vs)}$ ($2.4\ee{4}\unit{cm^2/(Vs)}$) and mean free paths of $l_{\mathrm{m}} \approx 0.4\unit{\um}$ ($0.3\unit{\um}$) at a carrier density of $1\ee{12}\unit{cm^{-2}}$ (Fig.~S1,~\cite{supplemental}).  The back-gate voltage is zero unless stated otherwise.

%
\begin{figure*}[t]
	\includegraphics[width=0.99\textwidth]{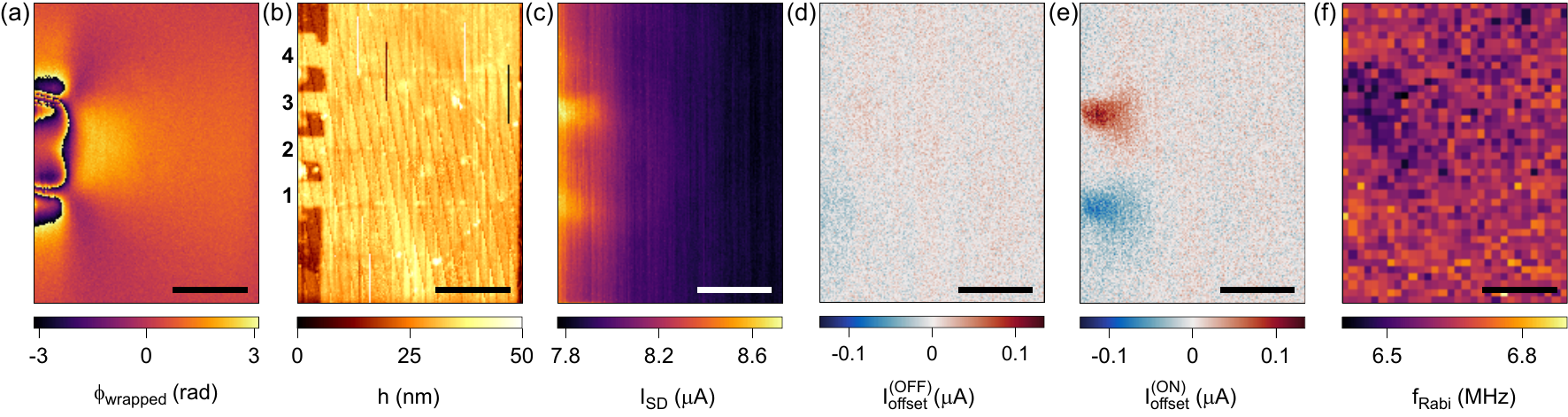}
	\caption{\captionstyle
	  {\bf Correlative maps of magnetometry and transport characteristics.}
		(a) Quantum phase recorded by the scanning magnetometer, according to Eqs.~(\ref{eq:counts}-\ref{eq:phiw}).
		(b) Topography recorded by the position feed-back of the scanning magnetometer.
		(c) AC amplitude of the source-drain current.
		(d,e) DC offset of the source-drain current while the laser is OFF (d) and ON (e) (see SI \cite{supplemental} for more details). Global offsets have been subtracted from both images.
		(f) Rabi frequency of the NV spin, recorded separately.
		Scale bars are $1\unit{\um}$.}
	\label{fig2}
\end{figure*}

Our custom-built scanning magnetometer consists of a three-axis sample stage that is scanned underneath a diamond probe tip in non-contact mode.  The diamond tip contains a single NV center near the apex.  The vertical stand-off between the NV center in the tip and the buried graphene sheet during a magnetometry scan is approximately $z=100\unit{nm}$ ($71\unit{nm}$ NV stand-off distance \cite{supplemental}, $11\unit{nm}$ hBN thickness, and $10-30\unit{\nm}$ additional scan distance).
A single diamond probe (count rate $C_0 \sim 550\unit{kC/s}$, spin contrast $\epsilon \sim 26\%$, QZabre Ltd. \cite{qzabre}) is used for all experiments.  Optical excitation (520\,nm) and detection (630-800\,nm) of the NV spin state are performed via the same objective located above the probe and sample.  Microwave excitation is achieved via a short bond wire loop ($\sim 30\unit{\um}$ away) that is not mechanically connected to the sample stage.   A small bias field of $5-18\unit{\mT}$ is applied to separate the NV $m_S=1\pm$ spin levels.  All measurements are carried out at room temperature.

\textit{Measurement technique -- }
We detect the current-generated magnetic field at each pixel using the concept of the quantum lock-in amplifier \cite{kotler11,delange11,ku20,vool21}.  We apply a square-wave voltage ($f = 50\unit{kHz} - 100\unit{kHz}$) between the source and drain contacts and synchronize the waveform with the microwave and laser pulses as well as optical detection.  To ensure proper synchronization of all channels, we generate all analog and digital signals on a multi-channel arbitrary waveform generator (AWG, Spectrum DN2.663-04), see Fig.~\ref{fig1}(d).  Another channel of the AWG is used to dynamically adjust the back-gate voltage during measurements.

The signal of the quantum lock-in is the quantum phase $\phi$ that the NV spin acquires during the coherent precession time $\tau$ (see Fig.~\ref{fig1}(d)).  For our protocol, the quantum phase is given by \cite{supplemental}:
\begin{align}
\phi = \ye\Bac\tau \ ,
\label{eq:phi}
\end{align}
where $\Bac$ is the signal amplitude and $\ye = 2\pi\times 28\unit{GHz/T}$ the gyromagnetic ratio of the NV electronic spin.  We determine $\phi$ via photo-luminescence (PL) intensity measurements,
\begin{align}
C_\Phi = C_0\left(1-\frac{\epsilon}{2} + \frac{\epsilon}{2}\cos(\phi + \Phi)\right)
\label{eq:counts}
\end{align}
where $C_\Phi$ is photons per second, $C_0$ is the photon count rate of the $m_S=0$ spin state, and $\epsilon$ is the optical contrast.  $\Phi$ is the relative phase of the final $\pi/2$ pulse.  By recording $C_\Phi$ for the read-out phases $\Phi = 0,\pi/2,\pi,3\pi/2$ (corresponding to the qubit axes $x$, $y$, $-x$, $-y$), we can determine the phase over the full $(-\pi;\pi]$ range using the two-argument arc-tangent \cite{knowles16,ku20}:
\begin{align}
\phiw = \tan ^{-1}\left(\frac{C_{-y} - C_{y}}{C_x - C_{-x}}\right)\ .
\label{eq:phiw}
\end{align}
This ``wrapped phase'' $\phiw$ is equal to $\phi$ modulo $2\pi$.  Therefore, for signals exceeding a maximum field $\Bmax = \pm \pi/(\ye\tau)$, we expect phase wrapping to occur in the image.  In this paper we develop suitable phase unwrapping techniques to recover the original phase $\phi$ with high dynamic range.

\textit{Correlative magnetometry and transport maps -- }
We begin our measurements by imaging the magnetic field from the four-terminal device (device A).  We apply a voltage of $\Vsd = 65\unit{mV}$ between contacts 1 and 3 (2 and 4 are floating), corresponding to a current of $\Isd \approx 8\unit{\uA}$, and use a spin echo sequence ($N=1, \tau= 10\unit{\us}$, see Fig.~\ref{fig1}(d)) to detect the ac modulation.  Fig.~\ref{fig2}(a) shows the wrapped quantum phase $\phiw$ extracted using the procedure of Eqs.~(\ref{eq:counts}) and (\ref{eq:phiw}).  Simultaneously with the phase measurement, we monitor the total PL signal (see \cite{supplemental}), device topography (Fig.~\ref{fig2}(b)), source-drain current amplitude $\Isd$ (Fig.~\ref{fig2}(c)) and dc offset (Figs.~\ref{fig2}(d,e)).  We further measure the Rabi frequency of the NV spin as a function of tip position in a separate scan (Fig.~\ref{fig2}(f)).

%
\begin{figure*}
	\includegraphics[width=0.75\textwidth]{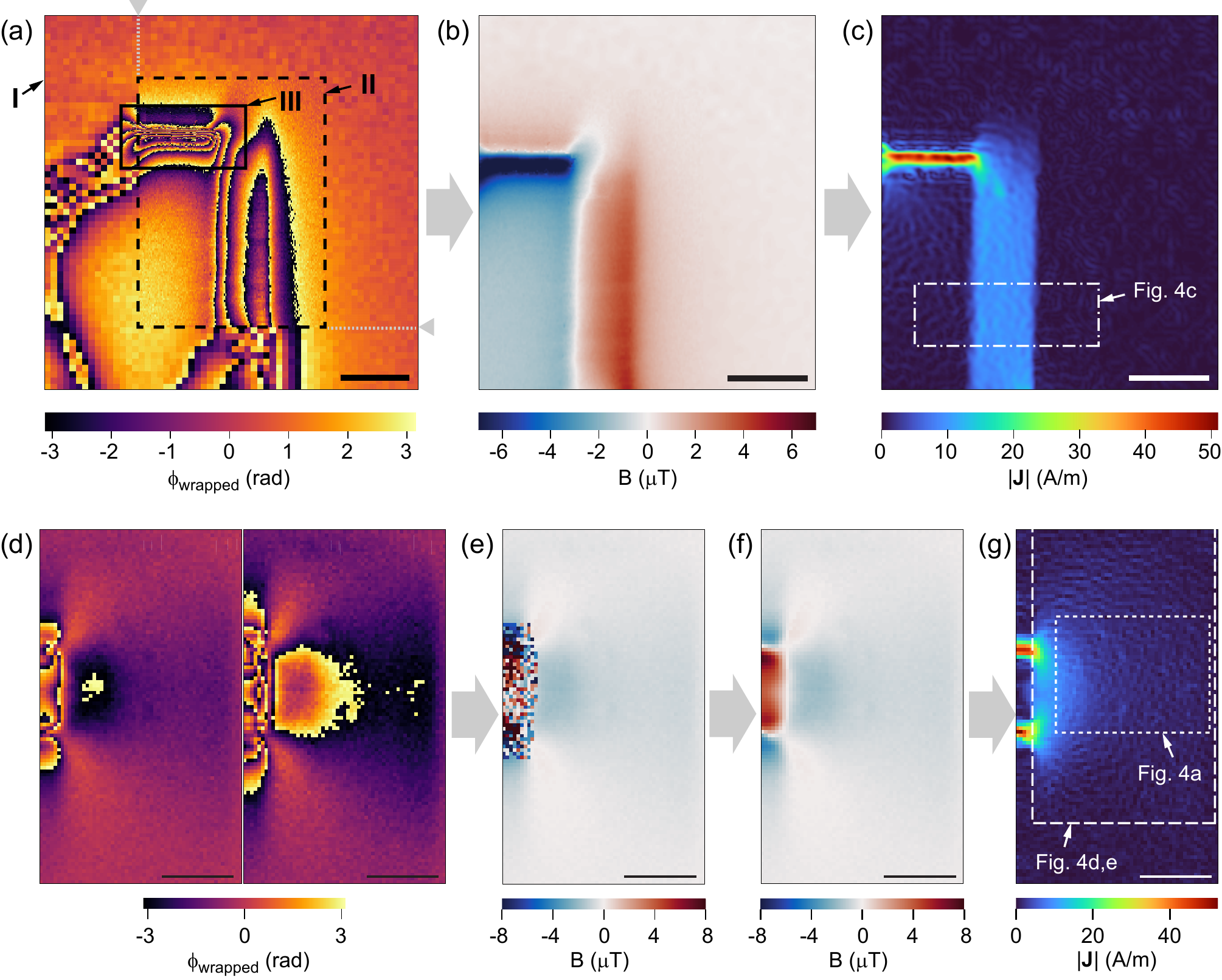}
	\caption{\captionstyle
	  {\bf Phase-unwrapping and current-density reconstruction.}
	  (a-c) Variable grid size method, demonstrated on device B with $\Isd \approx 7.2\uA$ between contacts 1 and 2.
		(a) Map of the quantum phase composed of three scans (I-III) with varying pixel sizes (I: $100\times 100\unit{nm}$, II: $19\times 33\unit{nm}$, III: $20\times 10\unit{nm}$).  Gray markers denote the sub-section of the image processed in (b-c).
		(b) Map of the magnetic field after applying a phase unwrapping algorithm. $B$ represents the vector field component along the NV center's anisotropy axis \cite{supplemental}.
		(c) Map of the current density reconstructed from (b).
		%
		(d-g) Bayesian inference method, demonstrated on device A with $\Isd \approx 10\uA$ between contacts 1 and 3.
		(d) Maps of the quantum phase recorded with $\tau_1=10\unit{\us}$ and $\tau_2=20.5\unit{\us}$.
		(e) Map of the magnetic field after the first phase-unwrapping step.
		(f) Map of the magnetic field after the second phase-unwrapping step that corrects for spatial smoothness.
		(g) Map of the current density.
		Dashed white boxes in (c,g) refer to Fig.~\ref{fig4}.
		Filter cut-off in (c,g) is $\lambda = z$, where $z$ is the stand-off distance.
		Scale bars are $1\unit{\um}$.
	}
	\label{fig3}
\end{figure*}

The correlative maps shown in Fig.~\ref{fig2}(b-e) allow us to monitor whether the tip, laser or microwave irradiation are affecting the transport properties of the device.
Proceeding from left to right, the current amplitude map $\Isd$ (Fig.~\ref{fig2}(c)) reveals a slight increase in the source-drain current when the tip is positioned near the injection or collection points, likely caused by a ``scanning gate'' effect due to trapped charges on the diamond tip or photo-doping (see below).  Other measurements on the same device show no tip influence or a reduction of the source-drain current with the tip positioned near a contact, showing that this effect is small but somewhat random.
Next, maps of the dc offset measured with the laser off and on (Figs.~\ref{fig2}(d) and (e), respectively, see Fig.~\ref{fig1}(d,e) for protocol) show that laser illumination can induce a small photo-current, especially when the tip is near the metallic contacts.  This photo-current effect is well-known \cite{lee08nnano, park09}, however, the effect is small and does not affect magnetometry because the quantum phase measurement always occurs in the laser off state.  In addition, to mitigate photo-doping of hBN, which causes drifts in the carrier density \cite{ju14, ku20, supplemental}, we ramp $\Vbg$ to zero during laser pulses (Fig.~\ref{fig1}(d)).
Fig.~\ref{fig2}(f) confirms that the Rabi frequency varies less than $5\%$ over the entire scan window, allowing us to rule out significant coupling of microwave pulses to the graphene device \cite{chang17,vool21}.  

\textit{Phase unwrapping -- }
We return to the magnetometry map shown in Fig.~\ref{fig2}(a).  Because of phase wrapping near the injection points where the current density is high, the phase map cannot be directly inverted to reveal the magnetic field $B$ and the associated current density $\vecJ$.  Therefore, in Fig.~\ref{fig3}, we develop two strategies to recover the magnetic field map even in the presence of large currents.

A first strategy is to use a variable grid and locally refine the pixel sizes in areas of rapidly changing field~\cite{jenkins20}.  Fig.~\ref{fig3}(a) displays experimental data taken on device B.  The phase map is composed of three separate scans (I-III) with pixel sizes between $10-100\unit{nm}$.  The pixel resolutions are chosen such that the true phase differences between pixels are roughly smaller than $\pi$.  We unwrap each phase map individually using a standard unwrapping algorithm \cite{phaseunwrap, herraez02}, convert the maps to units of magnetic field (Eq.~(\ref{eq:phi})), and interpolate them on a common $20\times 20\unit{nm^2}$ grid \cite{supplemental}.  The resulting field map is shown in Fig.~\ref{fig3}(b).  In a last step, we compute the current density map, Fig.~\ref{fig3}(c), by inverting Biot and Savart's law using an inverse filtering technique \cite{roth89,chang17,broadway20}.
The phase unwrapping increases the dynamic range of the image from $\Bmax = \pm \pi/(\ye\tau) \approx \pm 1 \uT$ to $\Bmax \approx \pm 6.5 \uT$, corresponding to a factor of $6.5\times$.

%
\begin{figure*}
	\includegraphics[width=0.90\textwidth]{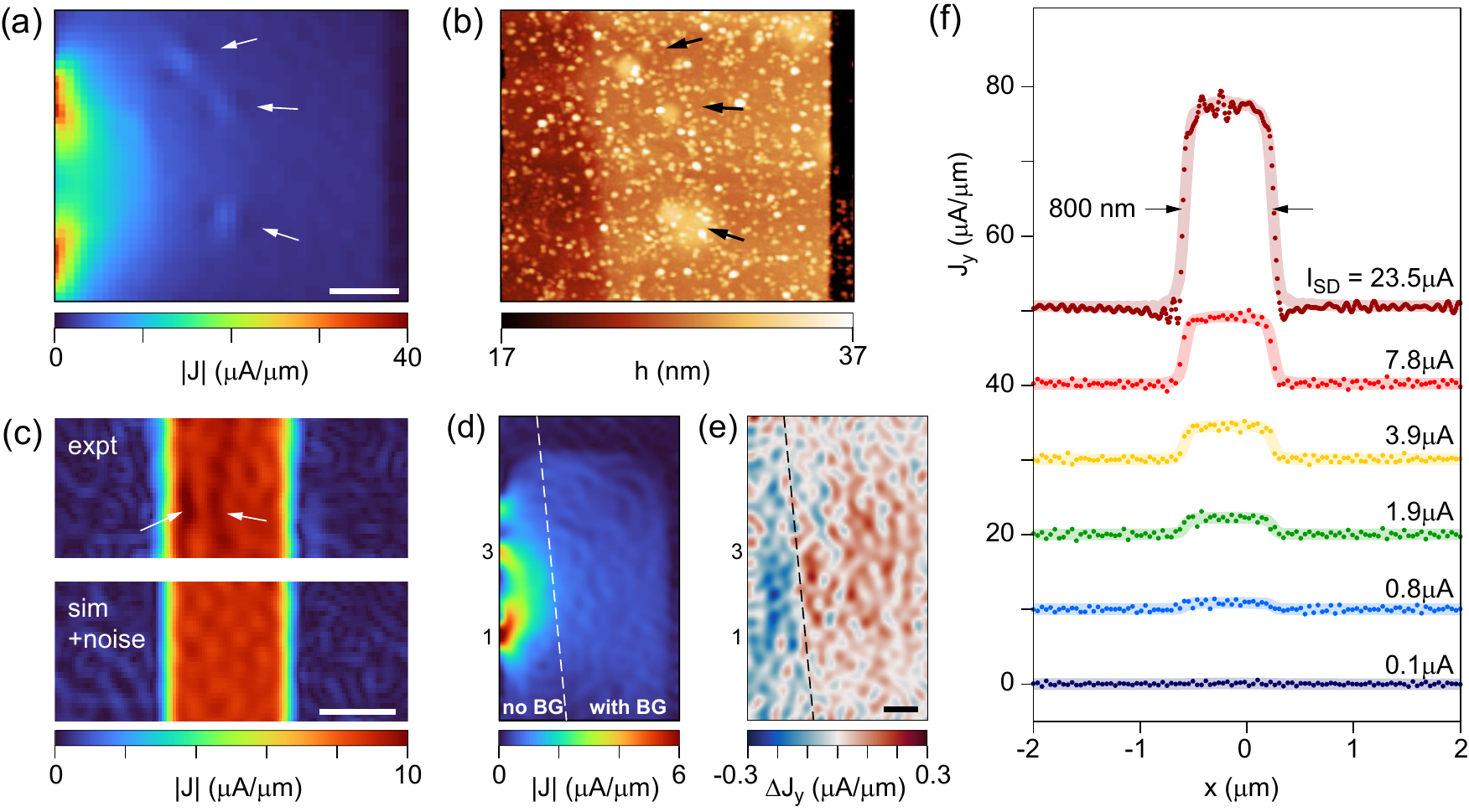}
	\caption{\captionstyle
	{\bf Transport physics}.
	(a) Magnified current density map of device A (dotted area in Fig.~\ref{fig3}(g)). Arrows indicate anomalies in the current flow.
	(b) Corresponding AFM topography.
	(c) Current density map of device B (dashed-dotted area in Fig.~\ref{fig3}(c)). Upper map shows the experimental data. Lower map shows simulated data assuming a perfect rectangular conductor after adding the equivalent amount of white noise. Arrows indicate channels of increased current flow in the experiment.
	(d) Current density map of device A (dashed area in Fig.~\ref{fig3}(g)) for $\Vbg=0$.  Note that the back-gate only covers the right part of the device, shown by a white contour.
	(e) Differential current density map $\Delta J_y$ obtained by subtracting $J_y^{(\Vbg=-2\mr{V})} - J_y^{(\Vbg=0\mr{V})}$.
	(f) Reconstructed current density profiles $J_y$ across the Hall-bar channel (device B).  Dots are the data and solid line is a calculation for a uniform current density profile. $\Isd$ is the applied source-drain current.  Step size is $10\unit{nm}$ for the $\Isd=23.5\uA$ curve and $40\unit{nm}$ for all other curves.  Curves are vertically offset for clarity.
	Scale bars are $500\unit{nm}$.
	}
	\label{fig4}
\end{figure*}

Our second approach to resolving the phase wrapping is based on a Bayesian inference and demonstrated in Fig.~\ref{fig3}(d-g) on device A.  We proceed in two steps: In a first step, we record two images with different interaction times $\tau_1=10\unit{\us}$ and $\tau_2=20.5\unit{\us}$, displayed in Fig.~\ref{fig3}(d).  The interaction times are chosen at high points of the spin echo curve where the sensitivity is maximum (Suppl. Fig. S10).  The global phase is then recovered either by inverse-variance weighting or by evaluating the joint probability function $P(B)$ \cite{nusran12, waldherr12} (see SI for details \cite{supplemental}).
Fig.~\ref{fig3}(e) shows the resulting field map.  Evidently, the first phase unwrapping step is not complete and fails in areas of high current density.  To improve the field estimation, in a second step, we invoke the fact that the magnetic field is spatially smooth and therefore, neighboring pixels are expected to have similar values.  According to Bayes' rule, the updated probability function is given by
$P(B|B_\mr{est}) \propto P_\mathrm{G}(B-B_\mr{est}) P(B)$,
where $P_\mathrm{G}(B-B_\mr{est})$ is a Gaussian centered around the weighted average of the neighboring pixels \cite{supplemental}.  The multiplication introduces an envelope to $P(B)$ that narrows down the set of possible probability maxima.  We then update $P(B)$ for all pixels by traversing the grid multiple times until convergence is achieved.  The results of this iterative phase estimation algorithm are presented in Figs. \ref{fig3}(f,g). 
Overall, we find that both phase-unwrapping methods are successful in recovering the field maps, however, the variable-grid method is somewhat unsatisfactory due to image stitching and possibility of image artifacts.

\textit{Transport physics -- }
We now turn our attention to the interpretation of the current density maps.  In particular, we inspect them for possible spatial signatures of non-uniform conductivity \cite{clark13, tetienne17, lillie19} and hydrodynamic transport \cite{levitov16,torre15,bandurin16,sulpizio19,jenkins20,ku20}.

Fig.~\ref{fig4}a shows a section of device A on a magnified scale.  While the current density is spatially smooth overall, we do observe channels of locally enhanced current flow (white arrows).   Although these anomalies are of order of a few $\uA/\um$ only, they are statistically significant and reproducible \cite{supplemental}.  Comparison with the AFM image (Fig.~\ref{fig4}(b)) reveals that the anomalies are correlated with slight rises in the  topography (black arrows). Similar statistically significant current density features are seen on device B (Fig.~\ref{fig4}(c)).  While we do not know the microscopic origin of the current density variations, it is likely that they reflect variations in conductivity due to a varying background potential \cite{pascher12,garcia13}.  Although encapsulating graphene in hexagonal boron nitride helps reducing the effect of charge impurities from the substrate \cite{dean10,xue11naturematerials}, the stacking process can lead to the formation of bubbles that act as dopants and local scatterers for transport \cite{leconte17}.  This explanation would be consistent with the correlated AFM topography observed with Fig.~\ref{fig4}(a,b).

To further investigate the influence of the local potential, we record current density maps for different values of the back-gate voltage akin to Ref.~\cite{lillie19}.  Since the expected current density variations are small, we use a differential acquisition technique where two images are recorded synchronously by toggling $\Vbg$ between subsequent measurement cycles (Fig. \ref{fig1}(d)).  The synchronous imaging guarantees that neither spatial drifts nor temporal changes in the transport properties result in spurious signals in the differential image.  Fig.~\ref{fig4}(d) shows a current density map for $\Vbg=0$ and Fig.~\ref{fig4}(e) the difference image between a $\Vbg=-2\unit{V}$ and $\Vbg=0\unit{V}$ map, respectively (see SI for the full data set).  These maps are recorded on device A where the graphite back-gate only covers part of the device (separated by the dashed line).  Consistent with a higher carrier density in the back-gated region (Suppl. Fig.~S1), the difference image is positive in the right portion of the map, confirming that current flow shifts to the high-conductivity region.  Overall, Fig.~\ref{fig4}(e) demonstrates that we can reliably detect small $\sim 5-10\%$ changes in the flow pattern despite the presence of a large background current density.

To reveal the possible presence of hydrodynamic transport effects, we analyze the current profile across the Hall-bar channel (device B).  A hallmark (but not unique \cite{sulpizio19}) signature for hydrodynamic transport is a parabolic flow profile, rather than a uniform (rectangular) profile associated with diffusive transport \cite{torre15, kiselev19}.  Recent experiments on monolayer graphene (MLG) have reported parabolic flow in channels of similar width at room temperature \cite{ku20}.  Fig. \ref{fig4}(f) shows a set of line scans across the Hall-bar channel for applied currents $0.1-23.5\unit{\uA}$.  All scans are taken near the charge neutrality point ($\Vbg=0$) where the carrier-carrier scattering is predicted to be strongest \cite{ho18}.
We find that all scans exhibit a rectangular current density profile (background lines in Fig. \ref{fig4}(f)) consistent with transport that is fully in the diffusive regime.  The absence of any hydrodynamic component could be due to a much lower carrier viscosity in BLG compared to MLG near charge neutrality for BLG \cite{bandurin16}.  Even higher mobilities or cryogenic temperatures may be needed to observe the effect, if at all present.

%
\begin{figure}
	\includegraphics[width=0.50\textwidth]{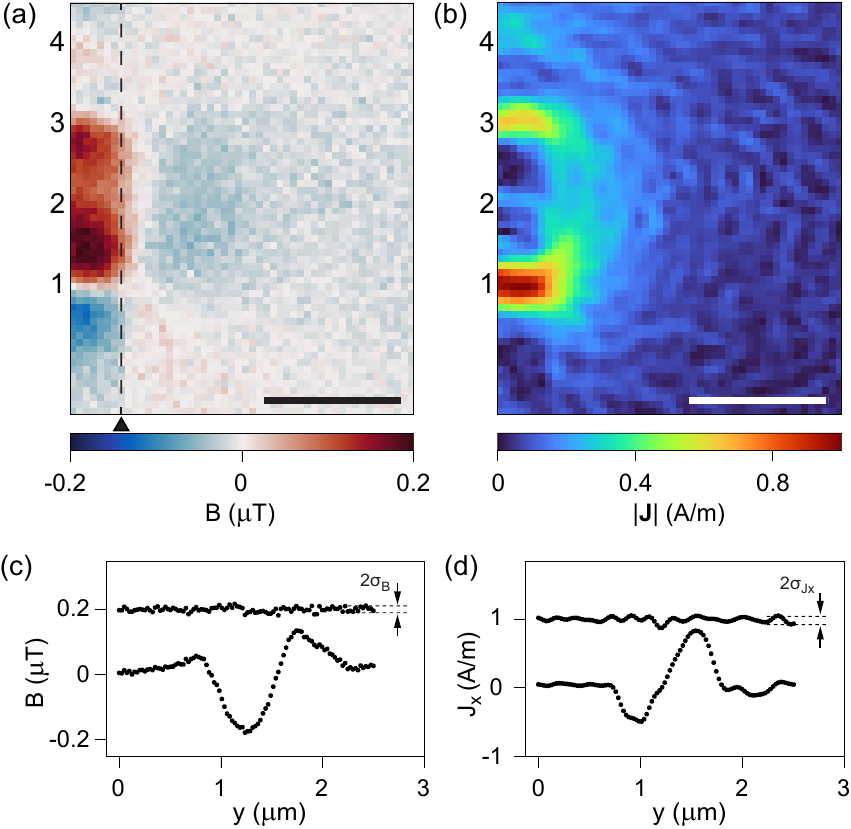}
	\caption{\captionstyle
	  {\bf Imaging of a $0.3\unit{\uA}$ current.}
		(a,b) High-sensitivity magnetic field and current density maps recorded on device A.  A current of $\Isd = 0.3\unit{\uA}$ and $f=1.33\unit{MHz}$ is injected into terminal 1 and collected at terminal 3.  Some current leakage through the floating terminals 2 and 4 is also observed.	Data are recorded using a dynamical decoupling sequence with $N=128$ refocusing pulses and $\tau = 48\unit{\us}$.  Scale bars are $1\unit{\um}$ and filter cut-off in (b) is $\lambda = 1.5z$.
		(c) Best-effort magnetic line scan along the dashed line in (a). 	For this scan, current is injected into terminal 2 and collected at terminal 1, and measurement parameters are $N=128$, $\tau=38\unit{\us}$ and $f=1.68\unit{MHz}$.  Each pixel represents a 120\,s average. The upper trace (vertically offset for clarity) shows the difference between two line scans. The standard deviation extracted from the point-to-point difference is $\sigma_B=4.6\unit{nT}$ \cite{supplemental}.
		(d) Corresponding current density line scan with a standard deviation of $\sigma_{J_x}=20\unit{nA/\um}$.
	}
	\label{fig5}
\end{figure}
%

\textit{High-sensitivity scans -- }
Finally, we explore the limits of our technique towards detection of small currents.  In Fig. \ref{fig5}(a,b), we perform two-dimensional imaging on device A with a source-drain current of $I=0.3\unit{\uA}$.  To maximize sensitivity, we increase the ac current modulation to $f=1.33\unit{MHz}$ and use a dynamical decoupling sequence with $N=128$ pulses to extend the interaction time to $\tau = 48\unit{\us}$ (Suppl. Fig. S10). 
From separate high-resolution line scans, shown in Fig. \ref{fig5}(c), we extract an absolute magnetic field sensitivity of $4.6\unit{nT}$ for an averaging time of $120\unit{s}$ per pixel (see Fig. S3 for full data). This corresponds to a per-root-Hertz sensitivity of $51\unit{nT/\rtHz}$, in good agreement with the nominal sensitivity expected for this NV center tip ($47\unit{nT/\rtHz}$, see \cite{supplemental}).
Likewise, from Fig. \ref{fig5}(d), we find an absolute current sensitivity of $20\unit{nA/\um}$ corresponding to a per-root-Hz sensitivity of approximately $0.2\unit{\uA/(\um\rtHz)}$.


\textit{Discussion -- }
In summary, we report on sensitive imaging of current flow in two-dimensional conductors using scanning diamond magnetometry.  We introduce a set of methods for increasing the sensitivity and dynamic range of the technique, for resolving small current density variations by synchronous differential imaging, and for mitigating undesired side-effects of magnetometry operation (due to, for example, the scanning tip, laser and microwave pulses) on the electronic transport.
These advances allow us to reveal subtle spatial variations in the current density in BLG devices, including anomalies resulting from bubbles in the hBN encapsulation and tuning of the flow pattern via the back-gate potential.  We also provide evidence that current flow is fully in the diffusive regime with no signs of carrier viscosity.

The sensitivity demonstrated in our work ($\sim 50\unit{nT/\rtHz}$) compares well to those reached with superconducting quantum interference devices mounted on scanning tips ($\sim 30\unit{nT/\rtHz}$, Ref.~\onlinecite{uri20nphys}).  The latter have recently allowed for impressive advances in the imaging of, for example, topological edge currents \cite{uri20nphys} or twist-angle disorder \cite{uri20nature}, but are confined to cryogenic temperatures.  
Looking forward, our techniques will therefore be especially useful for studying transport features over a wide temperature range, include hydrodynamic ``whirlpools'' \cite{guerrero19}, the graphene Tesla valve \cite{geurs20}, the onset of non-linearity in transport phenomena \cite{kouwenhoven89}, or the Stokes paradox in viscous two-dimensional fluids \cite{lucas17}.

%

\section*{Acknowledgments}

We thank M. Eich, A. Popert, F. de Vries, H. Overweg and A. Kurzmann for advice with the device fabrication, Z. Ding for support during preparatory experiments, and P. M\"arki and the FIRSTlab clean-room staff for technical support. 
This work was supported by the European Research Council through ERC CoG 817720 (IMAGINE), Swiss National Science Foundation (SNSF) Project Grant No. 200020\_175600, the National Center of Competence in Research in Quantum Science and Technology (NCCR QSIT), and the Advancing Science and TEchnology thRough dIamond Quantum Sensing (ASTERIQS) program, Grant No. 820394, of the European Commission. 
K.W. and T.T. acknowledge support from the Elemental Strategy Initiative conducted by the MEXT, Japan (Grant Number JPMXP0112101001) and JSPS KAKENHI (Grant Numbers 19H05790 and JP20H00354).

\textit{Author contributions -- }
C.L.D., M.L.P. and W.S.H. conceived the experiment.
M.L.P. prepared the bilayer-graphene sample with the help of P.R.
M.L.P. and W.S.H. implemented the ac magnetometry technique, carried out all magnetometry experiments and performed the data analysis.
M.L.P. performed the transport measurements.
P.W., S.E., P.S., S.D. and K.C. contributed to the hardware and software of the scanning magnetometer.
K.E. provided access to laboratory infrastructure and feedback on the manuscript.
M.L.P. and C.L.D. wrote the manuscript. All authors discussed the results.



\section*{Methods}

\textit{Device fabrication -- }
All flakes of the van-der-Waals stack are mechanically exfoliated onto silicon substrate chips with a $90\unit{\nm}$ oxide layer. Preselected flakes are subsequently picked up with a polymer stamp in a dry transfer process \cite{wang13,zomer14} in an argon atmosphere  (1. top hBN  2. (bilayer) graphene 3. bottom hBN 4. graphite) and deposited on a pre-patterned 3x3 mm$^2$ substrate chip. The remaining polymer residues are then dissolved in dichloromethane. The final stack is annealed at $350\unit{\degree C}$ for 3h in an argon atmosphere. We define electrical contacts in an e-beam lithography step using a bilayer of PMMA 50k (AR-P 630 series) and PMMA 950k (AR-P 670 series). We use an additional conductive polymer (AR PC 5090.02) as a top layer to mitigate charging during the e-beam exposure. To create a one-dimensional contact to the graphene sheet \cite{wang13}, we etch away sections of the top h-BN flake and partially the bottom h-BN flake (reactive ion etching with CHF$_3$/O$_2$).  The Cr/Au contacts are then deposited with an electron beam evaporator and the excessive metal is removed in a lift-off process. The process for patterning the device is very similar with the exception that the metal deposition step is omitted We contact the finished device with Al bond wires.

\textit{Experimental setup -- }
The scanning diamond magnetometer consists of a confocal microscope to read out the photo-luminescence of the NV center and an atomic force microscope to scan the sample with the diamond sensor. The diamond sensor is attached to a quartz tuning fork in an amplitude modulated shear-mode configuration \cite{qzabre}. The degeneracy of the $m_S=\pm 1$ states of the NV center is lifted by a bias field that is created by a movable permanent magnet beneath the sample holder. Microwave pulses are applied via a bond wire positioned close to the NV center. During a magnetometry scan, only the sample stage is moved (except for occasional optical re-alignment). The laser pulses are generated by a pulsed diode laser that was designed in-house. An arbitrary waveform generator (Spectrum DN2.663-04) synchronizes the laser pulses, the microwave pulses and the voltage signals sent to the graphene device. The device current is amplified with a transimpedance amplifier and recorded with the data acquisition module of a lock-in amplifier (Zurich Instruments MFLI). The photon signal of the NV center is captured by a single photon avalanche photo diode (Excelitas).

\textit{Diamond probe characterization -- }
We use the same diamond probe for all the experiments presented in this work. We typically start our experiments by aligning the external bias field to the NV's symmetry axis. We proceed by determining the resonance frequency for one of the two spin transition $m_S=0 \leftrightarrow m_S=\pm1$ through optically detected magnetic resonance (ODMR) spectroscopy. Next, we determine the durations for $\pi/2$ and $\pi$ pulses by measuring the Rabi oscillations as a function of microwave pulse length. To mitigate the effect of the $\sim 3.1\unit{\MHz}$ hyperfine splitting, we apply pulses exactly centered between the hyperfine peaks and aim for large Rabi frequencies, typically around $5-12\unit{\MHz}$.  Next, we record a spin echo or dynamical decoupling decay curve as detailed in Fig.~S10(a) to select $\tau$ values at maxima of the spin echo revivals \cite{childress06}.

Finally, we determine the stand-off distance $z$ between the NV center and the source of the magnetic signal (electrical current in our case) by scanning over the step edge of a thin film magnetic calibration sample (Pt/Co/AlOx). The expected magnetic field profile for this out-of-plane magnetized thin films is well understood \cite{hingant15} and can be fitted to the measured data.  For the scanning NV tip used throughout this study, we determine a mean standoff distance of $z = 71\unit{\nm}$, with a typical variation of $\pm 5\unit{\nm}$ between the 8 different line scans (see Fig.~S11).


%


%


%

\end{document}